\pdfoutput=1

\documentclass[11pt, a4paper]{article}

\usepackage{xcolor}
\usepackage{framed}
\usepackage{amsmath}
\usepackage{amsfonts}
\usepackage{amssymb}
\usepackage{graphicx, rotating}
\usepackage{epsfig}
\usepackage{latexsym}
\usepackage{graphicx}
\usepackage{color}
\usepackage{amsmath,bm,amssymb}
\usepackage{cite}
\usepackage{slashed}
\usepackage{hyperref}
\usepackage{epstopdf}

\usepackage{slashed}

\AppendGraphicsExtensions{.tif}
\hypersetup{colorlinks=true, citecolor=bluscuro, linkcolor=black, urlcolor=bluscuro}
\definecolor{rossos}{cmyk}{0,1,1,0.55}
\definecolor{bluscuro}{rgb}{0.15, 0.2, .85}
\definecolor{bluchiaro}{cmyk}{1,.3,0.,0.1}

\newcommand{\be}{\begin{eqnarray}}
\newcommand{\ee}{\end{eqnarray}}

\usepackage{amsmath}
\numberwithin{equation}{section}

\def\ba{\begin{eqnarray}}
\def\ea{\end{eqnarray}}

\def\bq{\begin{quote}}
\def\eq{\end{quote}}

 at 10truept

\newcommand{\beq}{\begin{equation}}
\newcommand{\eeq}{\end{equation}}
\newcommand{\beqa}{\begin{eqnarray}}
\newcommand{\eeqa}{\end{eqnarray}}
\newcommand{\bea}{\begin{eqnarray}}
\newcommand{\eea}{\end{eqnarray}}

\def\ltap{\ \raise.3ex\hbox{$<$\kern-.75em\lower1ex\hbox{$\sim$}}\ }
\def\gtap{\ \raise.3ex\hbox{$>$\kern-.75em\lower1ex\hbox{$\sim$}}\ }
\def\gl{\ \raise.5ex\hbox{$>$}\kern-.8em\lower.5ex\hbox{$<$}\ }
\def\roughly#1{\raise.3ex\hbox{$#1$\kern-.75em\lower1ex\hbox{$\sim$}}}

\def\del{\partial}

\def\d{{\rm d}}

 \def\N{{\cal N}}

\newcommand{\arXiv}[2]{\href{http://arxiv.org/pdf/#1}{{\tt [#2/#1]}}}
\newcommand{\arXivold}[1]{\href{http://arxiv.org/pdf/#1}{{\tt [#1]}}}

\allowdisplaybreaks

\begin{document}

\begin{titlepage}
\begin{flushright}
\end{flushright}
\vspace{0.1in}

\begin{center}
{\Large\bf\color{black} 
The Clockwork Supergravity
}\\
\bigskip\color{black}
\vspace{1cm}{
{\large Alex ~Kehagias$^{a}$ and Antonio ~Riotto$^b$}
\vspace{0.3cm}
} \\[7mm]
{\it {$^a$\, Physics Division, National Technical University of Athens\\ 15780 Zografou Campus, Athens, Greece}}\\
{\it $^b$ {Department of Theoretical Physics and Center for Astroparticle Physics (CAP)\\ 24 quai E. Ansermet, CH-1211 Geneva 4, Switzerland}}\\
\end{center}
\bigskip

\vspace{1cm}

\begin{abstract}
\noindent
We show  that the minimal $D=5$, $\N=2$ gauged supergravity set-up may encode naturally the recently proposed clockwork mechanism. The minimal embedding requires one vector multiplet in addition to the supergravity multiplet and the clockwork scalar is identified with the scalar in the 
vector multiplet. The scalar has a two-parameter potential and it can accommodate the clockwork, the Randall-Sundrum and  a no-scale  model with a flat potential, depending on the values of the parameters. The continuous clockwork background breaks half of the original supersymmetries, leaving a $D=4$, $\N=1$  theory on the boundaries. 
We also show that the generated hierarchy by the clockwork is not exponential but rather power law. The reason is that four-dimensional Planck scale has a power-law dependence on the compactification radius, whereas the corresponding KK spectrum depends on the logarithm of the latter.  

\end{abstract}
\bigskip

\end{titlepage}

\section{Introduction}
The clockwork   is an ingenious device which allows to start from a fundamental theory with no small fundamental parameters 
and obtain light degrees of freedom with suppressed interactions \cite{c1,c2,c3}.  Some applications of the clockwork mechanism
have been  worked out in a series of recent papers \cite{c4,c5,c6,c7,c8,c9,c10,c11,c12,c13,nilles,c14,c15}.
 
 The implementation of the clockwork  mechanism can be either  through a  discrete number of  fields or through the presence of an extra dimension, the so-called Continuous ClockWork (CCW).  The latter provides a possible solution to the  naturalness problem affecting the Higgs sector \cite{c3} and happens to be the  same as in linear dilaton duals of Little String Theory \cite{l1,l2}. More specifically, 
  a dilaton field $S$ is introduced within  a five-dimensional braneworld where the fifth dimension is  compactified on $S_1/{\mathbb Z}_2$. The  action is

\begin{eqnarray}
S&=&\int \d ^4x \d y\,\sqrt{-g}\left\{\frac{M_5^3}{2}
\left(R-\frac{1}{3}\partial_M S\partial^M S+4k^2 e^{-\frac{2}{3}S}\right)\right.\nonumber\\
&-&\left.\frac{e^{-\frac{1}{3}S}}{\sqrt{g_{55}}}
\left[\delta(y)\Lambda_0+\delta(y-\pi R)\Lambda_\pi\right]\right\}, \label{ccw}
\end{eqnarray}
where $R$ is the radius of the fifth dimension, $k^2$ parametrises the negative bulk vacuum energy,  $M_5$ is the fundamental scale in the bulk, and $\Lambda_0$ and $\Lambda_\pi$ are tensions on the brane satisfying the relation $\Lambda_0=-\Lambda_\pi=-4 k M_5^3$. The resulting  metric
is \cite{c3}

\beq
\d s^2=e^{\frac{4}{3}k|y|}\left(\eta_{mn}\d x^m\d x^n+\d y^2\right), \label{metccw}
\eeq
with $\eta_{mn}$ the flat Minkowski metric ($m,n=0,\cdots, 3$). Hierarchies are generated on the $y=\pi R$ brane with exponential suppressions of the form  $e^{-k\pi R}$ as in the Randall-Sundrum (RS) case \cite{RS1,RS2}.
In addition, the $y=y_0=\rm{const.}$ sections of the CCW metric (\ref{metccw}) are flat Minkowski spacetimes rescaled by the exponential factor 
$e^{\frac{4}{3}k|y_0|}$ again as in  RS. 
Hence, one is tempting to conclude that the generated hierarchy is exponential and therefore, with a $k R\sim 10$, a hierarchy of scales 
as large as $10^{13}$ can  easily be generated. We will see here that this is not  true and the generated hierarchy is only power law.  Indeed, in the RS case, $R$ denotes the compactification scale, i.e. the only physical scale besides the $D=5$ Planck mass $M_5$ and $k$. In the CCW case on the other hand, $R$ is not physical in the sense that it does not corresponds to any physical scale. The physical scale in the CCW is also the compactification radius which, as we will see, it is proportional to the hierarchy factor $e^{k\pi R}$. Therefore, the generated 
hierarchy is much more weak now and in particular the CCW generates only 
a power law hierarchy. 

We will show here that the CCW can be consistently embedded in the minimal 
${\cal N}=2$, $D=5$ gauged  supergravity \cite{GST0,GST1,GST-1,CDA,dWvP,sierra,lukas1,lukas2,gunaydin2,berg,ellis,ferrara0} with a vector multiplet. The latter contains a single scalar which is actually the CCW scalar $S$, with  a two-parameter  potential. When one of the two parameters of the scalar potential  vanishes, we get the CCW 
 and if the other parameter vanishes, the potential is flat. Finally, when both parameters are non-zero, a stable minimum exists with a negative cosmological constant corresponding to the known RS case. In all the above three case, half of the   supersymmetries are preserved, the corresponding branes are BPS and the theory on the 
two boundaries is just ${\cal N}=1$, $D=4$ supergravity.  In addition, 
we show that the CCW can also be embedded in extended $D=5$, ${\cal N}=4$  supergravity with $SU(2) \otimes U(1)$  as gauge group \cite{romans} and the CCW scalar is identified with the single scalar of the 
supergravity multiplet in this case. In addition, it turns out that the  parameter 
$k$ in the CCW action (\ref{ccw}) is just the gauge coupling of the $SU(2)$ group.

The paper is organised as follows: in  section 2 we show how the CCW can be derived from  the gauged $D=5$, $\N=2$ supergravity. In section 3 we construct the corresponding BPS states and include the presence of branes in section 4. Section 5 provides a possible M-theory embedding and section 6 
some generalisations of the previous results and some comments about the differences between the known RS-construction and the clockwork set-up.
Finally, we conclude in section 7.

\section{The continuous clockwork  from the gauged $D=5$, $\N=2$ supergravity}
Let us consider the minimal $D=5$, $\N=2$ supergravity coupled to $n_V$ vector multiplets \cite{GST0,GST1,GST-1}. Its extension containing additional $n_H$ hypermultiplets and $n_T$ tensor multiplets has been constructed in Ref. \cite{CDA}. The field content of the theory  is
\begin{eqnarray}
\left(e^{m}_\mu,\psi_\mu^i,A_\mu^I,\lambda^{i\, x},\phi^x\right) , \label{mult}
\end{eqnarray}
where  the gravitino $\psi^i_\mu$ and the gauginos $\lambda^{i\, x}$ $(i=1,2), ~(x=1,\cdots,n_V)$ are doublets under the $SU(2)_R$ $R$-symmetry,  $A_\mu^I$ ($I=0,1,\ldots, n_V)$ are the graviphoton and the vector of the vector multiplets and $\phi^x$ 
are the scalars of the vector multiplet. When a $U(1)$ subgroup of the full $SU_R(2)$ $R$-symmetry group is gauged, the bosonic part of the gauged  $D=5$ and $\N=2$ theory is 
\begin{eqnarray}
e^{-1}{\cal L}_{\rm bos}(k)
 &=&\frac{1}{2}R-\frac{1}{4}a_{IJ}F_{\mu\nu}^I
 F^{J\, \mu\nu}
 -\frac{1}{2}g_{xy}\del_\mu \phi^x\del^\mu \phi^y\nonumber \\
 &&+\frac{1}{6\sqrt{6}}C_{IJK}e^{-1}\epsilon^{\kappa\mu\nu\rho\sigma}F^I_{\kappa\mu}
 F^J_{\nu\rho}A^K_\sigma- g^2P(\phi). \label{act}
 \end{eqnarray} 
The scalar field target space is a very special
manifold, described by the cubic surface  
\begin{eqnarray}
 C_{IJK}h^I(\phi)h^J(\phi)h^K(\phi)=1, \label{hyper}
 \end{eqnarray}  
where the $n_V+1$ coordinates $h^I$ parametrise   the ambient space. 
Scalar target spaces as cosets of the Jordan family together with  their properties  
have been introduced and discussed in Refs. \cite{GST0,GST1,GST-1} and a 
complete classification for homogeneous scalar target spaces is given in 
Ref. \cite{dWvP}. 
%
%
 In addition, we have that 
\begin{eqnarray}
&& a_{IJ}=h_I h_J+h_{xI}h^x_J, ~~~h^I_x=-\sqrt{\frac{3}{2}}\partial_x h^I(\phi),\nonumber \\
 &&g_{xy}=h_x^Ih_y^Ja_{IJ}, ~~~ h_I=C_{IJK}h^J(\phi)h^K(\phi).
 \end{eqnarray} 
The supersymmetry transformation that leave invariant the $\N=2$ theory are 
\begin{eqnarray}
\delta e^m_\mu&=& \frac{1}{2}\overline{\epsilon}^i\gamma^m\psi_{\mu\, i},\nonumber \\
\delta \psi_{\mu\, i}&=&D_\mu(\widehat{\omega}) \epsilon_i+\frac{1}{4}\sqrt{\frac{1}{6}}i h_I 
\left({\gamma_\mu}^{\nu\rho}-4\delta_\mu^\nu \gamma^\rho\right)\widehat{F}_{\nu\rho}^I\epsilon_i+\frac{1}{48}
\gamma_{\mu\nu\rho} \epsilon^j\overline \lambda ^b_i\gamma^{\nu\rho}\lambda_j^b,\label{s1}\\
&&-\frac{1}{12}\gamma_{\mu\nu}  \epsilon^j\overline \lambda ^b_i\gamma^{\nu}\lambda_j^b-\frac{1}{12}\gamma^\nu  \epsilon^j\overline \lambda ^b_i\gamma_{\mu\nu}\lambda_j^b+\frac{1}{6} \epsilon^j\overline \lambda ^b_i\gamma^{\mu}\lambda_j^b+\frac{i\,g}{2}\sqrt{\frac{1}{6}} P_0(\phi)\gamma_\mu\delta^{ij}\epsilon_j, \label{s2}\\
\delta A_\mu ^I&=& -\frac{1}{2}h^I_x\overline{\epsilon}^i\gamma_\mu \lambda ^x_i+\frac{\sqrt{6}}{2}i \overline{\psi}^i_\mu \epsilon_i h^I,
\label{3} \\
\delta \lambda_i^x &=& -\frac{i}{2} \widehat{\slashed{\partial}}\phi^x-\frac{1}{2}
\overline{\epsilon}^j\lambda^z_j \lambda_i^y \Omega_{zy}^{x}+\frac{1}{4} h^x_I\gamma^{\mu\nu}\epsilon_i\widehat{F}_{\mu\nu}^I+\frac{1}{\sqrt{2}}g P^x(\phi) \delta^{ij}\epsilon_j\nonumber  \\
&& -\frac{1}{4\sqrt{6}}iT_{xyz} \Big{(}3 \epsilon^j\overline{\lambda}_i^y\lambda_j^z-\gamma_\mu \epsilon^j\overline{\lambda}_i^y\gamma^\mu \lambda_j^z-\frac{1}{2}
\gamma_{\mu\nu} \epsilon^j\overline{\lambda}_i^y\gamma^{\mu\nu} \lambda_j^z\Big{)},\label{s4}\\
\delta\phi^x&=&\frac{1}{2}i \overline{\epsilon}^i\lambda^x_i,\label{s5}
\end{eqnarray}
where hat indicates supercovariantization, $\Omega_{zy}^x$ is the Riemannian connection on the hypersurface (\ref{hyper}) and $g$ is the gauge coupling. The scalar functions  $P_0$ and $P_x$ are such that 
\begin{eqnarray}
P=-P_0^2+P_x P^x,
\end{eqnarray}
where
\begin{eqnarray}
P_0=2h^I V_I, ~~~~P_x=\sqrt{2}h^I_x V_I, \label{pp} 
\end{eqnarray}
and $V_I$ are $n_V+1$ arbitrary constants. 
 The quantities $a_{IJ},~h^I, ~h_x^I, T_{xyz}, P_0$,  and $P_x$ satisfy a number of constraints \cite{GST0,GST1,GST-1}. 
It is possible to determine the model once the data $C_{IJK}$ are
given. In particular, one may define $n_V+1$ real variables $X^I$ and define the prepotential ${\cal V}$ 
\begin{eqnarray}
  \mathcal{V}=\beta^3\, C_{IJK} X^IX^JX^k, 
  ~~~~\beta=\sqrt{\frac{2}{3}}.
  \end{eqnarray}  
  Then the matrix $a_{IJ}$ and $h_I$ are determined by 
  \begin{eqnarray}
  a_{IJ}&=&-\frac{1}{2}\frac{\partial^2}{\partial X^I\partial X^J}\ln N, \label{aij}\\
  h_I&=&\frac{1}{3 \beta}\frac{\partial}{\partial X^I} 
  \ln \mathcal{V}\Big|_{\mathcal{V}=1}. 
  \end{eqnarray}
 Here  we are interested in the simple case of   a single vector multiplet couplet to $D=5$, $\N=2$ supergravity. In this case, ($I,J=0,1$) and $X^I,~h^I$ are parametrised by a single scalar $\phi^x=\phi$ where $(x=1)$. 
We will further take 
\begin{eqnarray}
C_{011}= \frac{1}{3},  
\end{eqnarray}
as the only non vanishing component of $C_{IJK}$,
so that 
\begin{eqnarray}
\mathcal{V}&=&\beta^3  X^0 (X^1)^2,\nonumber\\
a_{ij}&=&{\rm diag}\left(\frac{1}{2(X^0)^2},\,\frac{1}{(X^1)^2}\right),\nonumber\\
h_I&=&\frac{1}{3\beta}\left(\frac{1}{X^0},\, \frac{2}{X^1}\right).
\end{eqnarray}
The $\mathcal{V}=1$ line is then determined by 
\begin{eqnarray}
\beta^3 \gamma^3 X^0 (X^1)^2=1,
\end{eqnarray}
which we parametrise as
\begin{eqnarray}
X^0=\frac{1}{\beta} e^{2b \phi},~~~~X^1=\frac{1}{\beta}
e^{-b\phi}.
\end{eqnarray}
The scalar target space metric turns then out then to be
\begin{eqnarray}
g_{xx}=h_{Ix}h_{Jx}a^{IJ}=\frac{4b^2}{3\beta^4}.
\end{eqnarray}
By demanding $g_{xx}=1$ so that the kinetic term of the scalar $\phi$ is canonically normalised, we find that 
\begin{eqnarray}
b=\pm\frac{1}{\sqrt{3}}. 
\end{eqnarray}
We will choose $b=-1/\sqrt{3}$ from now on.
Using that 
\begin{eqnarray}
h^I=\frac{2}{3\beta}\left(X^0,\, X^1\right),
\end{eqnarray}
we get from Eq. (\ref{pp}) that 

\begin{eqnarray}
 P_0&=&V_0 e^{-\frac{2\phi}{\sqrt{3}}}+V_1e^{\frac{\phi}{\sqrt{3}}}, \nonumber \\
 P_x&=&-V_0 e^{-\frac{2\phi}{\sqrt{3}}}+\frac{1}{2}V_1 e^{\frac{\phi}{\sqrt{3}}},
 \end{eqnarray} 
and
 \begin{eqnarray}
 P&=&-3 V_1\left(V_0 e^{-\frac{\phi}{\sqrt{3}}}+\frac{1}{4}V_1. 
 e^{\frac{2\phi}{\sqrt{3}}}\right).
 \end{eqnarray} 
The scalar potential turns out to be
\begin{eqnarray}
 V=g^2P=-3 g^2V_1\left(V_0 e^{-\frac{\phi}{\sqrt{3}}}+\frac{1}{4}V_1e^{\frac{2\phi}{\sqrt{3}}}\right),
 \end{eqnarray} 
 and the bosonic part of the Lagrangian for vanishing gauge fields
 turns out to be
\begin{eqnarray}
 e^{-1}{\cal L}_{\rm bos}&=&\frac{1}{2}R
 -\frac{1}{2}\del_\mu \phi\,\del^\mu \phi+3 V_1\left(V_0 e^{-\frac{\phi}{\sqrt{3}}}+\frac{1}{4}V_1
 e^{\frac{2\phi}{\sqrt{3}}}\right). \label{act01}
 \end{eqnarray} 
 The constants $V_0,V_1$ specify the theory and they can be chosen at will. For the case  $V_0=0$ we find that 
\begin{eqnarray}
 P_0=V_1\, e^{\frac{\phi}{\sqrt{3}}}, ~~~
 P_x=\frac{1}{2}V_1 e^{\frac{\phi}{\sqrt{3}}}, 
 \end{eqnarray}
 so that the scalar  potential is written as
 \begin{eqnarray}
 V&=&-2 k^2 e^{\frac{2\phi}{\sqrt{3}}},
 \end{eqnarray} 
 where we have defined the parameter $k$ as 
 \begin{eqnarray}
 k=\sqrt{\frac{3}{8}}\, g V_1. \label{kg}
 \end{eqnarray}
 Hence,   Eq. (\ref{act01}) is written in this case 
 \begin{eqnarray}
 e^{-1}{\cal L}_{\rm bos}(k)&=&\frac{1}{2}R
 -\frac{1}{2}\del_\mu \phi\,\del^\mu \phi+2k^2 e^{\frac{2\phi}{\sqrt{3}}}. \label{act1}
 \end{eqnarray} 
The Lagrangian (\ref{act1}) can be written in many equivalent ways and appeared previously in the literature. For example, it coincides with a non-critical string theory in five dimensions. 
Indeed, by defining the dilaton $\Phi$ and the string metric 
$g^{(\sigma)}_{\mu\nu}$ as 
\begin{eqnarray}
\Phi=-\frac{\sqrt{3}}{2}\phi, ~~~g^{(\sigma)}_{\mu\nu}=e^{-\frac{2}{3}\Phi}g_{\mu\nu}, 
\end{eqnarray}
we may write (\ref{act1}) as 
\begin{eqnarray}
 e^{-1}{\cal L}_{\rm bosonic}^{\N=2}&=&\frac{1}{2}e^{-2\Phi}\Big(R
 +4\del_\mu \Phi\,\del^\mu \Phi+\delta c \Big), \label{act10}
 \end{eqnarray}
 where  
\begin{eqnarray}
\delta c=2k^2,
\end{eqnarray}
is the central charge deficit. For a string propagating in $D$-dimensions with $D<D_{\rm crit}$, $\delta c$ is 
\begin{eqnarray}
\delta c=\frac{2(D_{\rm crit}-D)}{3\alpha'},
\end{eqnarray}
 and (\ref{act10}) describes the low-energy non-critical string effective action. Sub-critical string theories  ($\delta c>0$)  could be  the result of tachyon condensation as discussed in \cite{pol}.
In addition, it has employed in Ref.  \cite{AADG} as a gravity dual of Little String Theory. What is more important, it  is also the action that describes the bulk of the $D=5$  CCW \cite{c3} after defining  
\begin{eqnarray}
\phi=-\frac{S}{\sqrt{3}}. \label{S}
\end{eqnarray}
Indeed, in this case Eq. (\ref{act1}) is written as 
\begin{eqnarray}
e^{-1}{\cal L}_{\rm bos}(k)&=&\frac{1}{2}\Big(R
 -\frac{1}{3}\del_\mu S\,\del^\mu S+4k^2 e^{-\frac{2S}{3}}\Big), \label{act2}
\end{eqnarray}
which is precisely the continuous CCW action.

Finally, let us also mentioned that the Lagrangian (\ref{act01}) appears also in $D=5$ extended supergravity \cite{romans,at,gian}. In particular, the $SU(2)\otimes U(1)$ gauged supergravity multiplet of the $D=5$, ${\cal N}=4$ theory is 
\begin{eqnarray}
\left(e^{m}_\mu,\psi_\mu^i,A_\mu^I,a_\mu, B_{\mu\nu}^a,\chi^{i},
\phi\right) , \label{mult}
\end{eqnarray}
i.e., it contains
the graviton $e^m_\mu$, four gravitini $\psi^i$ transforming in the 
fundamental representation  of the $USp(4)$ R-symmetry group,  four vector fields $A_\mu^I,a_\mu$  
of the $SU(2)\otimes U(1)$ gauge group, two antisymmetric gauge fields $B_{\mu\nu}^a$,  four spin-1/2
fermions $\chi^i$ also in the fundamental of $USp(4)$ and one real scalar field $\phi$. It turns out that the bosonic part of theory for vanishing gauge fields is 
\begin{eqnarray}
 e^{-1}{\cal L}_{\rm bos}^{\mathcal{N}=4}&=&\frac{1}{2}R
 -\frac{1}{2}\del_\mu \phi\,\del^\mu \phi+g_S\left(g_A e^{-\frac{\phi}{\sqrt{3}}}+g_S 
 e^{\frac{2\phi}{\sqrt{3}}}\right), \label{act00}
 \end{eqnarray} 
where $g_S,g_A$ are the gauge couplings of the $SU(2)$ and $U(1)$ gauge fields respectively \cite{romans,gian}. A simple inspection of (\ref{act01}) and (\ref{act00}) reveals that the constants $V_0,V_1$ are related to the gauge couplings $g_S,g_A$. In particular, the CCW Lagrangian (\ref{act1}) corresponds to 
\begin{eqnarray}
g_A=0,~~~g_S=\sqrt{2}\, k,
\end{eqnarray}
so that the CCW parameter $k$ is identified in the ${\cal N}=4$ context with the $SU(2)$ gauge coupling.  The $g_A=0$ in the full theory seems to be singular \cite{romans} but as it has been shown in \cite{con} it can be consistently be taken after appropriate field redefinitions.

\section{The continuous clockwork and supersymmetric backgrounds}
The Lagrangian in Eq. (\ref{act}) is invariant under eight (real) supersymmetries generated by the symplectic Majorana spinors $\epsilon^i$. We will look for solutions here to the field equations that preserves some supersymmetry \cite{wall-1,wall-2,BKvP,BKvP2,CDA2}. 
The  field equations turn out to be
\begin{eqnarray}
&&R_{\mu\nu}-\frac{1}{2}g_{\mu\nu} R=\del_\mu \phi\del_\nu \phi-\frac{1}{2}\Big{(}
(\del\phi)^2-4 k^2 e^{\frac{2\phi}{\sqrt{3}}}\Big{)}, \label{ein}\\
&&\nabla^2\phi+\frac{4}{\sqrt{3}}k^2 e^{\frac{2\phi}{\sqrt{3}}}=0. \label{ein2}
\end{eqnarray}
In addition, the fermionic shifts for the $n_V=1$ model are 
\begin{eqnarray}
\delta \psi_{\mu }^i&=&D_\mu \epsilon^i+\frac{k}{3}i \gamma_\mu  e^{\frac{\phi}{\sqrt{3}}}\delta^{ij}\epsilon_j,\\
\delta \lambda^{i}&=& -\frac{i}{2} \gamma^\mu\partial_\mu\phi\epsilon^i+\frac{k}{\sqrt{3}} 
e^{\frac{\phi}{\sqrt{3}}} \delta^{ij}\epsilon_j. \label{susy}
\end{eqnarray}
We look for backgrounds with four-dimensional Poincar\'e symmetry, which we can express as 
\begin{eqnarray}
{\rm d}s^2=e^{2\sigma(y)}\Big{(}\eta_{mn}{\rm d}x^m{\rm d}x^n+{\rm d}y^2\Big{)}, 
\end{eqnarray}
 Using the fact that for conformally related metrics $\widetilde{g}_{\mu\nu}=\Omega^2g_{\mu\nu}$ we have that 
\begin{eqnarray}
 \widetilde{D}_\mu\epsilon=D_\mu \epsilon+\frac{1}{2}{\gamma_\mu}^\nu(\partial_\nu \ln \Omega)\,\epsilon,
 \end{eqnarray} 
we find that the condition for unbroken supersymmetry (vanishing fermionic shifts) is
explicitly written as 
\begin{eqnarray}
&&\Big{(}\delta_k^i \gamma_5 \sigma'+\frac{2k}{3}ie^{\frac{\phi}{\sqrt{3}}+\sigma}\delta^{ij} \varepsilon_{jk}\Big{)}\epsilon^k=0, \label{s1}\\
&& {\epsilon^{i}}'+\frac{k}{3}ie^{\frac{\phi}{\sqrt{3}}+\sigma}\gamma_5\delta^{ij} \epsilon_j=0, \label{s2}\\
&&\Big{(}\delta_k^i \gamma^5\phi'-\frac{2k}{\sqrt{3}}ie^{\frac{\phi}{\sqrt{3}}+\sigma}\delta^{ij} \varepsilon_{jk}\Big{)}\epsilon^k=0, \label{s3}
\end{eqnarray}
 where $\varepsilon_{ij}$ is the $SU(2)$ invariant antisymmetric tensor and a prime denotes differentiation with respect to $y$. Then, Eq. (\ref{s2}) determines the dependence of the Killing spinor on the $y$-coordinate, whereas Eqs. (\ref{s1}) and (\ref{s3}) as  projector equations require the consistency conditions
 \begin{eqnarray}
&& \left(\delta^i_j {\sigma'}^2-\frac{4k^2}{9} e^{\frac{\phi}{\sqrt{3}}+\sigma}\varepsilon^{in}\varepsilon_{nj}\right)\epsilon^j=0, \label{c1}\\
&& \left(\delta^i_j {\phi'}^2-\frac{4k^2}{3} e^{\frac{\phi}{\sqrt{3}}+\sigma}\varepsilon^{in}\varepsilon_{nj}\right)\epsilon^j=0, \label{c2}
 \end{eqnarray}
and lead to 
\begin{eqnarray}
\sigma'=\pm \frac{2k}{3}e^{\frac{\phi}{\sqrt{3}}+\sigma}, ~~~~\phi'=\mp \frac{2k}{\sqrt{3}}e^{\frac{\phi}{\sqrt{3}}+\sigma}. \label{ss}
\end{eqnarray}
Solving Eq. (\ref{ss})  we find 
\begin{eqnarray}
\sigma=\frac{-\phi}{\sqrt{3}}=\frac{2k}{3}y, \label{sol}
\end{eqnarray}
for increasing scale factor and conditions $\sigma(0)=\phi(0)=0$. 
It is straightforward to check that the field equations Eqs. (\ref{ein}) and (\ref{ein2}) are satisfied as expected. Then  the background is 
\begin{eqnarray}
{\rm d}s^2=e^{\frac{4k}{3}y}\Big{(}\eta_{mn}{\rm d}x^m{\rm d}x^n+{\rm d}y^2\Big{)}, ~~~S=2ky, \label{back}
\end{eqnarray}
where $S=-\sqrt{3}\, \phi$ is the CCW scalar. This is the linear dilaton solution employed in \cite{AADG,c3}. 
The spacetime described by the 
metric (\ref{back}) has a singularity at $y\to \infty$. In the next section, we will introduce branes at finite values of $y$  so that the spacetime is non-singular with boundaries.

It should be noted that this configuration is a BPS state. Indeed for static configurations we may define the energy functional  for the CCW background

\begin{eqnarray}
E=-S_{\rm bosonic}=e^{3\sigma}\Big(6{\sigma'}^2-\frac{1}{6}
{S'}^2+2k^2e^{2\sigma-\frac{2S}{3}}\Big)-4e^{3\sigma}\sigma' \Big|_{-\infty}^{\infty}.
\end{eqnarray}
Note that since 
\begin{eqnarray}
2k^2e^{-\frac{2S}{3}}=g^2(-P_0^2+P_x^2),
\end{eqnarray}
we can write the energy functional as 
\begin{eqnarray}
E=e^{3\sigma}\Big\{ (\sqrt{6}{\sigma'}-g P_0e^{\sigma})
(\sqrt{6}{\sigma'}+g P_0e^{\sigma})-(\frac{1}{\sqrt{6}}
{S'}+g P_xe^{\sigma})(\frac{1}{\sqrt{6}}
{S'}-g P_xe^{\sigma})\Big\}-4e^{3\sigma}\sigma' \Big|_{-\infty}^{\infty}.
\end{eqnarray}
The structure of the energy suggests the BPS condition
\begin{eqnarray}
\sigma'=\pm \frac{g}{\sqrt{6}}P_0e^{\sigma}, ~~~S'=\mp \sqrt{6} g P_x e^{\sigma}.  \label{bps}
\end{eqnarray}
For such BPS configurations, the value of the energy turns out to be  $E=0$, in accordance with the result of Ref.  \cite{BKvP,BKvP2}. 
It is easy to verify that the BPS condition above is identical to the supersymmetry preserving condition (\ref{ss}). 

\section{Branes in the continuous clockwork }
Let us now assume that there is one direction, let say the $x^5=y$ direction which parametrise the circle $S^1/\mathbb{Z}_2$. The $\mathbb{Z}_2$ orbifolding introduces two fixed points at $y=0$ and $y=\pi R$ where $r$ is the $S^1$ radius. These are singular points, but we can give a physical meaning to them by assuming that they are the location of $D=4$ branes. Bulk fields have definite $\mathbb{Z}_2$ parity and they can be either even ($\Phi(-y)=\Phi(y)$), or odd ($ \Phi(-y)=-\Phi(y)$)
 under the $y\to -y$ inversion. Therefore, odd field at the orbifold fixed
 points where the branes  reside vanish and should not appear in the $D=4$ theory on the branes. Similarly to the bulk fields, also the supersymmetry parameters are split accordingly into half even ($\epsilon_+$) and half odd ($\epsilon_-$) spinors. Clearly, the brane theory would be invariant only under the $\epsilon_+$, i.e., only under an ${\cal N}=1$ supersymmetry.

In order to write a supersymmetric theory in this setup, one has to assume that the coupling $g$ is not continuous but it has jumps at the brane positions. In other words, we assume that the coupling is $g$ for $0<y<\pi R$ whereas it is $-g$ for $-\pi R<y<0$. Thus, we introduce the function 
$\hat{g}(y)$ as 
\begin{eqnarray}
   \hat g(y)=g, ~~~0<y<\pi R, ~~~~~\hat g(y)=-g, ~~~ -\pi R<y<0, 
   \end{eqnarray}   
   which satisfies
   \begin{eqnarray}
   \del_y \hat g=2 g\Big(\delta(y)-\delta(y-\pi R)\Big).
   \end{eqnarray}
   In addition, due to Eq. (\ref{kg}), we will can define $\hat k=\sqrt{8/3} \hat g V_1$, which satisfies similarly
   \begin{eqnarray}
   \del_y \hat k=2 k\Big(\delta(y)-\delta(y-\pi R)\Big).
   \end{eqnarray}
The bosonic part of the supersymmetric Lagrangian on the theory is 

\begin{align*}
e^{-1}{\cal L}_{\rm new}&= e^{-1}{\cal L}_{\rm bos}(\hat k)+
  A_{\mu\nu\rho\sigma} \epsilon^{\mu\nu\rho\sigma\kappa}\partial_\kappa\hat k
  \nonumber \\
  &-2k \Big(\delta(y)-\delta(y-\pi R)\Big)\Big(2e^{-1}e_{(4)}e^{\phi/\sqrt{3}} + \frac{1}{2}\epsilon^{mnpq}A_{mnpq}\Big),
\end{align*}
  where $e_{(4)}$ is the determinant of the induced vielbein on the branes and $m,n,...=0,1,2,3$.
  Then the action of the theory is written as 
\begin{eqnarray}
S=\int_{M_4\times S^1/\mathbb{Z}_2}\, \d^4x\d y\,
  {\cal L}_{\rm  new}, \label{Snew}
  \end{eqnarray}  
  is supersymmetric. Indeed, (\ref{Snew})  
 is invariant under the supersymmetry transformations $\hat \delta (\epsilon)$, which are the transformations Eqs. (\ref{s1})-(\ref{s5}) with the coupling $g$ replaced with $\hat g$ and 
 \begin{eqnarray}
 \hat \delta A_{mnpq}&=&-\frac{1}{\sqrt{6}}\overline \epsilon _i 
 \gamma_{[mnp}\psi^i_{q]} P_0+i\overline \epsilon_i V_I\gamma_{[mn}A^I_{p}\psi^i_{q]}+\frac{i}{\sqrt{2}} \overline \epsilon_i \gamma_{mnpq}\lambda^i_x
 P^x,\nonumber \\
 \hat \delta \hat k&=&0. 
 \end{eqnarray}
 Now the condition on backgrounds that preserve some supersymmetry is the vanishing of the fermionic shifts which gives 
\begin{eqnarray}
&&\Big{(}\delta_k^i \gamma_5 \sigma'+\frac{2\hat k}{3}ie^{\frac{\phi}{\sqrt{3}}+\sigma}\delta^{ij} \varepsilon_{jk}\Big{)}\epsilon^k=0, \label{s12}\\
&& {\epsilon^{i'}}+\frac{\hat k}{3}ie^{\frac{\phi}{\sqrt{3}}+\sigma}\gamma_5\delta^{ij} \epsilon_j=0, \label{s22}\\
&&\Big{(}\delta_k^i \gamma^5\phi'-\frac{2\hat k}{\sqrt{3}}ie^{\frac{\phi}{\sqrt{3}}+\sigma}\delta^{ij} \varepsilon_{jk}\Big{)}\epsilon^k=0. \label{s32}
\end{eqnarray}
 The above equations have a solution if the consistency condition 
 \begin{eqnarray}
&& \left(\delta^i_j {\sigma'}^2-\frac{4k^2}{9} e^{\frac{\phi}{\sqrt{3}}+\sigma}\varepsilon^{in}\varepsilon_{nj}\right)\epsilon^j=0, \label{c12}\\
&& \left(\delta^i_j {\phi'}^2-\frac{4k^2}{3} e^{\frac{\phi}{\sqrt{3}}+\sigma}\varepsilon^{in}\varepsilon_{nj}\right)\epsilon^j=0, \label{c22}
 \end{eqnarray}
 (since $\hat k ^2=k^2$) is satisfied. 
 Hence, we find that in order to have some  unbroken supersymmetry, the scalars $\sigma$ and $\phi$ should obey
\begin{eqnarray}
\sigma'=\pm \frac{2k}{3}e^{\frac{\phi}{\sqrt{3}}+\sigma}, ~~~~\phi'=\mp \frac{2k}{\sqrt{3}}e^{\frac{\phi}{\sqrt{3}}+\sigma}. \label{ss2}
\end{eqnarray}
We may choose the branch $\sigma=-\phi/\sqrt{3}$, and then from Eq. (\ref{s12}) and Eq. (\ref{s32}) we find 
\begin{eqnarray}
\sigma=\frac{-\phi}{\sqrt{3}}=\frac{2k}{3}|y|, \label{sol2}
\end{eqnarray}
for $\sigma(0)=\phi(0)=0$.  Therefore   the background is 
\begin{eqnarray}
\d s^2=e^{\frac{4k}{3}|y|}\Big{(}\eta_{mn}\d x^m\d x^n+\d y^2\Big{)}, ~~~S=2k|y|, \label{back}
\end{eqnarray}
where $S=-\sqrt{3}\, \phi$ is the CCW scalar. Note that the this background breaks half of the supersymmetries. Indeed, from Eq. (\ref{s12}) for example, we get that $\epsilon^i$ should satisfy
\begin{eqnarray}
(1+\gamma^5)\epsilon^i=0\, 
\end{eqnarray}
  which means that out of the 8 real component $\epsilon^i$ only 4 real components are non zero, and therefore, the 
  boundary branes beaks half of the supersymmetries. The remaining 4 real component $\epsilon^i$ form a complete spinor in $D=4$ and preserve ${\cal N}=1$
  local supersymmetry on the $D=4$ boundary branes.  The spectrum of the boundary theory can easily be found.


The low-energy four-dimensional theory is obtained by the dimensional reduction of the five-dimensional fields. In particular, the five dimensional gravity and vector  multiplets
\begin{eqnarray}
 (e^a_\mu,\psi_\mu^i,A_\mu), ~~~(B_\mu,\lambda^i,\phi),
 \end{eqnarray} 
will split as 
\begin{eqnarray}
(e^a_m,e^5_m, e^5_5,\psi_m^L,\psi_m^R,A_m,A_5),~~~(B_m,B_5,\lambda_L,\lambda_R,\phi). \label{ff}
\end{eqnarray}
These fields fill the graviton and two vector massless  multiplets of the  four-dimensional ${\cal N}=2$ theory as 
\begin{eqnarray}
\overbrace{
\underbrace{\left(\begin{tabular}{c}
$g_{mn}$\\
$\psi_m^R$
\end{tabular}
\right)  }_\text{${\cal N}=1$ graviton}
\underbrace{\left(\begin{tabular}{c}
$g_{m5}$\\
$\psi_m^L$
\end{tabular}
\right)}_\text{${\cal N}=1$ gravitino}
}^{{\cal N}=2~ {\rm graviton}}, ~~~
\overbrace{
\underbrace{\left(\begin{tabular}{c}
$A_{m}$\\
$\psi_5^L$
\end{tabular}
\right)  }_\text{${\cal N}=1$ vector}
\underbrace{\left(\begin{tabular}{c}
$g_{55}, A_5$\\
$\psi_5^R$
\end{tabular}
\right)}_\text{${\cal N}=1$ chiral}
}^{ {\cal N}=2~ {\rm vector ~multiplet}},
~~~\overbrace{
\underbrace{\left(\begin{tabular}{c}
$B_{m}$\\
$\lambda^L$
\end{tabular}
\right)  }_\text{${\cal N}=1$ vector}
\underbrace{\left(\begin{tabular}{c}
$\phi,~B_5$\\
$\lambda^R$
\end{tabular}
\right)}_\text{${\cal N}=1$ chiral}
}^{ {\cal N}=2~ {\rm vector ~multiplet}}. \label{f1}
\end{eqnarray}
In (\ref{f1}) above, we express the ${\cal N}=2$ multiplets in terms of their ${\cal N}=1$ content.
The linear dilaton background beaks the ${\cal N}=2$ to ${\cal N}=1$ supersymmetry and the fields (\ref{ff}) are arranged in representations of the unbroken ${\cal N}=1$ supersymmetry. In particular, we will have a massless  ${\cal N}=1$ graviton, the ${\cal N}=1$ gravitino will eat the vector multiplet and will become massive and similarly the vector in the last multiplet will eat the chiral and become massive as well.   Therefore, the four-dimensional ${\cal N}=1$ spectrum will contains a massless graviton  multiplet $(2,3/2)$, a massive gravitino multiplet 
$(3/2,1,1,1/2)$ and a massive vector multiplet $(1,1/2,1/2,0)$ preserving the original $12_B+12_F$ degrees of freedom.  

\section{M-Theory Embedding}
Let us recall some well-known results of the compactification of M-theory on a Calabi-Yau three-fold $CY_3$ \cite{ferrara,ferrara2,ferrara3}, parametrised by complex coordinates $z^i$ ($i=1,2,3$) and with Hodge numbers $h_{(1,1)}$ and $h_{(2,1)}$ and intersection numbers $C_{IJK}$, where $(I,J,K=1,\cdots, h_{(1,1)}$. The $D=5$ bosonic degrees of freedom will come from the bosonic fields of 11D supergravity, namely, the metric $g_{MN}$ and the totally antisymmetric three-form $A_{MNK}$. After compactification on $CY_3$, the $D=5$ spectrum consists of a complex axion $A_{ijk}=\epsilon_{ijk} \chi$, a real axion 
dual to the fields strength of the three-form $A_{\mu\nu\rho}$, the CY volume $\det g_{i\overline j}$, $h_{(1,1)}-1$ real 
K\"ahler moduli $g_{i\overline j}$, $h_{(2,1)}$ complex scalars $(g_{ij}, A_{ij\overline k})$ and $h_{(1,1)}$ abelian vectors $A_{\mu i\overline j}$. These fields are paired with fermions to form supermultiplets of the $D=5$,
 ${\cal N}=2$ supergravity in the following way. One of the $h_{(1,1)}$ vectors is the graviphoton which together with the graviton $g_{\mu\nu}$ form the supergravity multiplet. The remaining vectors together with the scalars $g_{i\overline j}$ form  
$h_{(1,1)}-1$  vector multiplets. The fields $(\det g_{i\overline j}, 
a,A_{ijk})$ form the universal hypermultiplet and finally the fields ($g_{i\overline j},A_{ij\overline k})$ form $h_{(2,1)}$ hypermultiplets. We are interested here for the scalars in the vector multiplets which are just the 
K\"ahler moduli $t^I$. The latter can be defined after expanding the K\"hler form $J$ on the $CY_3$ into the basis of the fundamental 2-forms $J_I$ as
\begin{eqnarray}
  J=t^I J_I.
  \end{eqnarray}  
This definition of $t^I$ specifies them as the size of the 2-cycles 
of the $CY_3$. 
The $D=5$, ${\cal N}=2$ vector couplings are determined  by first
introducing  the superpotential ${\cal V}$ as 
\begin{eqnarray}
 \mathcal{V}=C_{IJK} t^It^Jt^K,
 \end{eqnarray} 
 which defines the very special geometry \cite{dWvP}.
Then, the independent scalar degrees of freedom are determined by the constraint 
\begin{eqnarray}
 \mathcal{V}=1,
 \end{eqnarray} 
 and the K\"ahler moduli can be identified with the functions $h^I$. In general, the scalar moduli space is of the form
 \begin{eqnarray}
 {\cal M}_V=SO(1,1)\otimes \frac{SO(1,n-1)}{SO(n-1)}, ~~~n=h_{(1,1)}-1.
 \end{eqnarray}
 With $n=2$, which is the case we are considering here, the moduli space is simply $SO(1,1)$ parametrised by the single scalar $\phi$. A simple model with $n=2$ is the two-parameter model with prepotential in the large volume limit
 
\begin{eqnarray}
\mathcal{V}=9(t^1)^3+9(t^1)^2(t^2)+3 (t^1) t^2)^2.
\end{eqnarray}
 Defining 
 \begin{eqnarray}
 h^1=t^1, (h^2)^2=9(t^1)^2+9(t^1)(t^2)+3 (t^2)^2,
 \end{eqnarray}
 we have 
 \begin{eqnarray}
 \mathcal{V}=h^1(h^2)^2, 
 \end{eqnarray}
 which is the model we considered in section 2.

 \section{Generalisations and some general considerations}
  Let us now generalise our previous findings and assume  that  both $V_0$ and  $V_1$ in  section 2 are non-vanishing. In this case the functions $P_0$ and $P_1$ are written in terms of the $S$ field defined in Eq. (\ref{S}) as 
 \begin{eqnarray}
 P_0&=&V_0 e^{\frac{2S}{3}}+V_1 e^{-\frac{S}{3}}, \nonumber\\
 P_x&=&-V_0 e^{\frac{2S}{3}}+\frac{1}{2}V_1 e^{-\frac{S}{3}}, \label{PP}
 \end{eqnarray}
  whereas the potential turns out to be 
  \begin{eqnarray}
 V=-2k^2\left(4V_0 e^{\frac{S}{3}}+V_1 
 e^{-\frac{2S}{3}}\right), ~~~~k=\sqrt{\frac{3}{8}}g V_1.\label{vvv}
 \end{eqnarray} 
 Then, in terms of a new coordinate $r$ defined by 
 \begin{eqnarray}
 \d r=e^{\sigma(y)}\d y,
 \end{eqnarray}
  the BPS conditions (\ref{bps}) turn out to be 
  \begin{eqnarray}
  \frac{\d\sigma}{\d r}=\pm \frac{g}{\sqrt{6}} P_0, ~~~\frac{\d S}{\d r}=
  \mp \sqrt{6} g P_x,
  \end{eqnarray}
 and they are explicitly written as
  \begin{eqnarray}
  \frac{\d\sigma}{\d r}&=&\pm \frac{g}{\sqrt{6}}\Big(V_0 e^{\frac{2S}{3}}+V_1 e^{-\frac{S}{3}}\Big),\\
  \frac{\d S}{\d r}&=&\mp \sqrt{6} g\Big(-V_0 e^{\frac{2S}{3}}+\frac{1}{2}V_1 e^{-\frac{S}{3}}
  \Big). \label{s10}
  \end{eqnarray}
  There are three distinct cases, depending on the values of the parameters $V_0$ and $V_1$: 

\begin{itemize}

\item $V_0=0$: This is the case  we discussed  above and corresponds to the linear dilaton case and  to the CCW. 
 
 \item  $V_1=0$:  For this choice of the parameters,  $k=0$, the potential $V=0$ and the theory is of no-scale type \cite{no-scale1,no-scale2,ferrara0}. Eq. (\ref{s10}) is 
  written in this case as
  \begin{eqnarray}
  \frac{\d\sigma}{\d r}&=&\pm \frac{g}{\sqrt{6}}V_0 e^{\frac{2S}{3}},\nonumber\\
  •\frac{\d S}{\d r}&=&\pm  \sqrt{6} gV_0 e^{\frac{2S}{3}}.\label{ssq}
  \end{eqnarray}
  The solution to Eqs. (\ref{ssq}) that satisfies also Einstein equations is in this case 
  \begin{eqnarray}
  \sigma&=&\frac{1}{4}\ln \Big(C_0+ \gamma\, r\Big)+\sigma_0\\
  S&=&\pm \frac{3}{2}\ln\Big(C_0+ \gamma\, r\Big), ~~~\gamma=\sqrt{\frac{8}{3}}g V_0,
  \end{eqnarray}
  where $C_0$ and $\sigma_0$ are   constants.  
Then the metric turns out to be 
\begin{eqnarray}
  \d s^2=r^{1/2}\, \eta_{mn}\d x^m
\d x^n+\d r^2,
\end{eqnarray}  
after appropriate redefinition and the coordinates 
and takes the conformaly flat form 
\begin{eqnarray}
\d s^2=(1+\mu y)^{2/3} \Big(\eta_{mn}\d x^m
\d x^n+\d y^2\Big), 
\end{eqnarray}
where $\mu$ a mass scale. 
There is a singularity at $r=0$ (or $y=0$). This singularity is harmless   if we assume that $y$ parametrise $S^1/\mathbb{Z}_2$ with 
$y_0=0\leq y\leq y_\pi=\pi R$. In this case we find the background metric 
\begin{eqnarray}
  \d s^2=\Big(1+\mu |y|\Big)^{2/3}\, \Big(\eta_{mn}\d x^m
\d x^n+\d y^2\Big),  ~~~~~\mu>0,
\end{eqnarray} 
  which is supported by an energy-momentum tensor of the form
  \begin{eqnarray}
  T_{mn}=\eta_{mn}\Big(\Lambda_0 \delta(y-y_0)+\Lambda_\pi \delta(y-y_\pi)\Big), ~~~T_{m5}=T_{55}=0,
  \end{eqnarray}
   where 
   \begin{eqnarray}
   \Lambda_0=-\frac{2\mu }{M_5^3}, ~~~\Lambda_\pi=\frac{1}{M_5^3}
   \frac{2\mu }{1+\mu \pi R}.
   \end{eqnarray}
  This is the energy-momentum tensor of a planar branes sitting at $y=y_0$ and $y=y_\pi$ with corresponding tensions $\Lambda_0$ and $\Lambda_\pi$.

\item Both $V_0$ and $V_1$ different from zero:   In this case, for $V_0,V_1>0$, we find that there is a local maximum of the potential corresponding to the AdS$_5$ or RS-background with cosmological constant $\Lambda= -4k^2\sqrt{V_0V_1}$. 
The background is written in conformal coordinates as 
  \begin{eqnarray}
  \d s^2=\frac{1}{(1+\mu|y|)^2}\Big(\eta_{mn}\d x^m \d x^n+\d y^2\Big),
  \end{eqnarray} 
  where $\mu=\sqrt{-\Lambda/6}.  $
  \end{itemize}
  
Let us now deconstruct the $S^1/\mathbb{Z}_2$ direction for the simple model of an scalar and a $U(1)$ theory on a background of the form
\begin{eqnarray}
\d s^2=a^2(y)\Big(\eta_{mn}\d x^m \d x^n+\d y^2\Big). 
\end{eqnarray}
and with action 
\begin{eqnarray}
S&=&-\frac{1}{2}\int \d^4x \int_{y_0}^{y_\pi} \d y
\Big\{ a^3(y)(\partial_\mu \phi)^2+\frac{1}{2}a(y) F_{\mu\nu}^2\Big\}=
\nonumber \\
&=&-\frac{1}{2}\int \d^4x \int_{y_0}^{y_\pi} \d y\Big\{(\partial_m \phi)^2+\frac{1}{2}F_{mn}^2+
a^3(\partial_y a^{-3/2}\phi)^2+a(\partial_y a^{-1/2} A_m)\Big\},
\end{eqnarray}
where we have assumed Dirichlet conditions for $A_y$ and 
$a(y)=(1+\mu |y|)^{1/3}$. After discretising the $y$ direction with $y=j {\rm a},~y_0=0,~y_\pi=N{\rm a}$, we may write the action as 
\begin{eqnarray}
{\cal S}&=&-\frac{1}{2}\int \d^4x \bigg\{\sum_{j=0}^N\Big[(\partial_m \phi_j)^2+\frac{1}{2}F_{jmn}^2\Big]+\nonumber \\
&&\frac{1}{\rm a}\sum_{j=0}^{N-1}\left\{ \left[\phi_j-
q_{j+1}\phi_{j+1}\right]^2+\left[A_{nj}-
q_{j+1}^{1/3}A_{n,j+1}\right]^2\right\}, \label{dec}
\end{eqnarray}
where 
\begin{eqnarray}
 q_j=\left(\frac{a(j)}{a(j+1)}\right)^{3/2}. \label{qjj}
\end{eqnarray}
This is the form of a discrete CW 
 with site-dependent deformation parameter $q$. Indeed, for the model with 
 $V_1=0$ we find that 
\begin{eqnarray}
 q_{j+1}=
  \left(\frac{j\mu y_\pi + N}{(j+1)\mu y_\pi+ N}\right)^{3/2}.
  \end{eqnarray}  
  Similarly,
 deconstructing the $S^1/\mathbb{Z}_2$ direction $y$ for the RS, we arrive at the same Eq. (\ref{dec}) with 
  \begin{eqnarray}
  q_{j+1}=\left(\frac{(j+1)\mu y_\pi+N}{j\mu y_\pi + N}\right)^3.
  \end{eqnarray}  
   Therefore, only the linear dilaton background when deconstructed gives a side-independent $q$ for flat boundary branes \cite{c3,c4}. In fact it can be proved that this is the only case with  a site-independent $q$-factor. 
Indeed, solving the recurrence Eq. (\ref{qjj}) with initial condition
  $a(0)=1$ and constant $q$ ($q_0=q_1=\cdots,q_N=q$), we find 
\begin{eqnarray}
  a(j)=q^{-\frac{2}{3}j}.
  \end{eqnarray}  
  In the continuum limit, we get then 
  \begin{eqnarray}
   a(y)=e^{\frac{2}{3} k y}, 
   \end{eqnarray} 
   where $k=(\log\, q)/3$, i.e. the CCW metric. In other words, only the CCW has a site independent q-factor. However,
  site-dependent $q$ has appeared also in CCW with curved boundary branes \cite{c4,nilles}.

 Let us now calculate the $D=4$ Planck mass $M_{P}$
  
  \begin{eqnarray}
  M_{\rm pl}^2=2M_5^3\int_0^{y_\pi} a^3(y) \d y=\left\{\begin{array}{ll}
  \mbox{$\frac{M_5^3}{k}\left(e^{2ky_\pi}-1\right)$}&\mbox{$V_0=0$ ~~~(CCW-linear dilaton),~~~}
\\
  \mbox{$M_5^3y_\pi(\mu y_\pi+2)$}&\mbox{$V_1=0$~~~({\rm no-scale})},~~~\\
  \mbox {$\frac{M_5^3}{\mu} \frac{2+\mu y_{\pi}}{1+\mu y_\pi}$}& \mbox{$V_0\neq 0$ and $V_1\neq 0$~~~ (RS)}.
   \end{array}\right.
  \end{eqnarray}
  The compactification radius $r_c$ turns out to be
  \begin{eqnarray}
  r_c=\int_0^{y_\pi} a(y) \d y=\left\{\begin{array}{ll}
  \mbox{$\frac{3}{2k}\left(e^{\frac{2ky_\pi}{3}}-1\right)$}& \mbox{$V_0=0$ ~~~(CCW-linear dilaton),~~~}
\\
  \mbox{$\frac{3}{4\mu}\Big\{(1+\mu y_\pi)^{4/3}-1\Big\}$}&\mbox{$V_1=0$~~~({\rm no-scale})},~~~
  \\
  \mbox {$\frac{1}{\mu }\ln(1+\mu y_\pi)$}&\mbox{$V_0\neq 0$ and $V_1\neq 0$~~~ (RS)}.
   \end{array}\right.
  \end{eqnarray}
Then, in terms of the physical $r_c$ scale, the $D=4$ Planck mass is written as
  
  \begin{eqnarray}
  M_{\rm pl}^2=\left\{\begin{array}{ll}
  \mbox{$\frac{M_5^3}{k}\Big\{(1+\frac{2k}{3}r_c)^{3}-1\Big\}$}& \mbox{$V_0=0$ ~~~(CCW-linear dilaton),~~~}
\\
  \mbox{$\frac{M_5^3}{\mu}\Big\{(1+\frac{4\mu}{3} r_c)^{3/2}-1\Big)$}&\mbox{$V_1=0$~~~({\rm no-scale})},~~~\\
  \mbox {$\frac{M_5^3}{\mu} \Big(1-e^{-\mu r_c}\Big)$}& \mbox{$V_0\neq 0$ and $V_1\neq 0$~~~ (RS)}.
   \end{array}\right.
  \end{eqnarray}
If expressed in terms of the physical radius $r_c$ only in the  RS compactification  the planckian mass has an exponential  dependence. In all other cases, and above all in the CCW, there is only a power-law dependence. This is analogous to what happens in inflationary models, where an exact  exponential expansion in the scale factor is obtained if the energy density of the universe is dominated exactly by a cosmological constant, while a power-law, but still accelerated,  expansion can be obtained as long as the equation of state has an index smaller than $-1/3$.

Therefore, we see that only in the case of the RS we have an exponential dependence of the $D=4$ Planck mass on the compactification radius. In the other two cases, the dependence is just power law \cite{kehagias}. However, it is interesting that although we have one extra dimension and the $D=4$ Planck mass would expected to depend only linearly on the compactification radius, in the case of the CCW a much stronger cubic dependance (as if we have two additional hidden dimension) and a fractional dependence in the complementary case.  In addition, for values of $M_5$ and $k$ 
\begin{eqnarray}
M_5\sim 10 ~{\rm TeV}, ~~~k\sim 1~{\rm TeV},
\end{eqnarray}
the compactification radius turns out to be
\begin{eqnarray}
r_c=1~ {\rm GeV}^{-1},
\end{eqnarray}
which is about a tenth of the proton size. 
In the limit of large compactification radius $r_c$,  the $D=4$ Planck mass scales as

  \begin{eqnarray}
  M_{\rm pl}^2\approx\left\{\begin{array}{ll}
  \mbox{$\frac{M_5^3}{k}\Big(\frac{2k}{3}\Big)^{3}\, r_c^{3}$}& \mbox{$V_0=0$ ~~~(CCW-linear dilaton),~~~}
\\
  \mbox{$\frac{M_5^3}{\mu}\Big(\frac{4\mu}{3}\Big)^{3/2}\, r_c^{3/2}$}&\mbox{$V_1=0$~~~({\rm no-scale})},\\
  \mbox {$\frac{M_5^3}{\mu} $}& \mbox{$V_0\neq 0$ and $V_1\neq 0$~~~ (RS)}.
   \end{array}\right.
  \end{eqnarray}
  In the other limit $kr_c\ll 1$ or $\mu r_c\ll 1$ all models satisfy the 
  usual $M_{\rm pl}^2\approx M_5^3 r_c$ relation. 
The dependence of the masses  KK states on $r_c$ is peculiar for the CCW. A  massless scalar $\phi$  obeys the equation 
\begin{eqnarray}
\Box \phi=0, \label{min}
\end{eqnarray}
which 
on the  CCW background (\ref{back}) is explicitly written as
\begin{eqnarray}
\del_m\del^m \phi+3\sigma'\del_y \phi+\del^2_y\phi=0.
\end{eqnarray}
We may express $\phi$ as $\phi(x,y)=e^{ip_m x^m }e^{-\frac{3}{2}\sigma}\psi(y)$ where $\psi$ satisfies 
\begin{eqnarray}
\psi''-\left(\frac{9}{4} (\sigma')^2+\frac{3}{2}\sigma''\right)\psi
=p^2 \psi, 
\end{eqnarray}
and therefore, the KK states are the eigenvalues of the 
\begin{eqnarray}
\psi_n''-\Big\{k^2+2k \delta(y)-2k\delta(y-y_\pi)\Big\}\psi_n=-m_n^2 \psi_n.
\label{psin}
\end{eqnarray}
Then even eigenfunctions $\psi_n$ satisfy the boundary condition
\begin{eqnarray}
\psi_n'-k\psi\big|_{y=0,y_\pi}=0.
\end{eqnarray}
The boundary condition at $y=0$ leads to 
\begin{eqnarray}
\psi_n=\frac{1}{N_n}\left\{\cos(\sqrt{m_n^2-k^2}|y|)-\frac{k}{\sqrt{m_n^2-k^2}}\sin(\sqrt{m_n^2-k^2}|y|)\right\},
\end{eqnarray}
whereas the condition at $y=y_\pi$ gives 
\begin{eqnarray}
\sin(\sqrt{m_n^2-k^2}y_p)=0,
\end{eqnarray}
from where the KK spectrum \cite{AADG}
\begin{eqnarray}
m_n^2=\frac{n^2\pi^2}{y_\pi^2}+k^2,
\end{eqnarray}
follows. The same spectrum is also found for odd eigenfunctions with Dirichlet boundary conditions. In terms of the compactification scale, the KK masses are expressed as
\begin{eqnarray}
m_n^2=\frac{4k^2 \pi^2 n^2}{9\log(1+2kr_c/3)^2}+k^2.
\end{eqnarray}
Note that in the $kr_c\ll 1$ limit we get the usual scaling of the KK masses with $r_c$
\begin{eqnarray}
m_n^2\approx \frac{\pi^2n^2}{r_c^2}+k^2,
\end{eqnarray}
whereas in the opposite limit $kr_c\gg 1$ we get 
\begin{eqnarray}
m_n^2=\frac{4k^2 \pi^2 n^2}{9\log(2kr_c/3)^2}+k^2.
\end{eqnarray}
The  shift in the KK mass spectrum by $k^2$ is due to the minimal coupling of the scalar to gravity. For a conformally coupled scalar in five dimensions,
\begin{eqnarray}
\Box \phi-\frac{3}{16} R\phi=0, 
\end{eqnarray}
 we find that on the CCW background where 
\begin{eqnarray}
R=-16  e^{-2\sigma}/3 k\Big(k+2k\delta(y)-k \delta(y-y_\pi)\Big),
 \end{eqnarray} 
Eq. (\ref{psin}) is written simply as 
\begin{eqnarray}
\psi_n''
=-m_n^2 \psi_n,
\end{eqnarray}
and the KK spectrum  turns out then to be (with Neumann boundary condition)
 \begin{eqnarray}
m_n=\frac{2\pi k \, n}{3\log(1+2kr_c/3)}.
\end{eqnarray}

 \section{Conclusions}

We have studied the CCW model in a $D=5$, ${\cal N}=2$ supergravity framework. The minimal embedding requires one $D=5$ vector multiplet, the scalar of which is the CCW scalar. After gauging the $SU(2)$ $R$-symmetry of the 
${\cal N}=2$ theory, a two-parameter potential for the scalar emerges.

This potential accommodates three-models, the RS model when both parameters are non-vanishing, the CCW when one parameter vanishes and a third model when the second parameter vanishes and leads in fact to  a vanishing potential. For the CCW in particular, we have shown that it preserves half of the supersymmetries and therefore, the $D=4$ effective theory is
$D=4$, ${\cal N}=1$  supergravity. We have also shown that, when the $D=5$ theory has $D=4$ boundary branes, these branes do not break completely supersymmetry,  but for the CCW background only half of the supersymmetries are broken. Hence, again the boundary theory is ${\cal N}=1$ $D=4$ supergravity.

We have calculated the compactification radius of these models and when the $D=4$ Planck mass is expressed in terms of the compactification radius, it follows that only the RS model has an exponential dependence on the latter. The other two models, the CCW and the third model have a power law dependence on the compactification radius. However, in the case of the CCW, the $D=4$ Planck mass dependence  is quite strong  on the compactification radius in the sense that it scales with the cube of it and not linearly as it would be expected for one extra dimension. Thus, it seems that the CCW looks as if there were three extra dimensions. We have also calculated the masses of the KK states for a scalar in CCW background and we found that they depend on the inverse of the logarithm of the compactification scale.

\section*{Acknowledgments}
We thank G. Giudice and M. McCullough  for reading the draft of the paper, for discussions and  for providing  useful  suggestions. A.K. is partially supported by GGET
project 71644/28.4.16. A.R. is supported by the Swiss National Science Foundation (SNSF), project {\sl Investigating the
Nature of Dark Matter}, project number: 200020-159223.

\end{document}